\documentstyle[preprint,aps,prb]{revtex}

\begin{document}

\draft
%\flushbottom
%\twocolumn[\hsize\textwidth\columnwidth\hsize\csname @twocolumnfalse\endcsname
%
\title{ First-Principles Wannier Functions of Silicon and Gallium Arsenide}
\author{Pablo Fern\'andez$^1$, Andrea Dal Corso$^1$, Francesco Mauri$^2$, and
Alfonso Baldereschi$^1$}
\address{
$^1$ Institut Romand de Recherche Num\'erique en Physique des Mat\'eriaux
(IRRMA), IN Ecublens, 1015 Lausanne, Switzerland. \\
$^2$ Department of Physics, University of California at Berkeley, Berkeley, CA
94720, USA,
and Materials Science Division, Lawrence Berkeley National Laboratory,
Berkeley, CA 94720, USA.
}
\date{\today}
\maketitle

\begin{abstract}

We present a self-consistent, real-space calculation of
the Wannier functions of Si and GaAs within density functional theory.
We minimize the total energy functional with respect to orbitals which
behave as Wannier functions under crystal translations and, at the minimum, are
orthogonal.
The Wannier functions are used to calculate the total energy, lattice constant,
bulk modulus, and the frequency of the zone-center TO phonon
of the two semiconductors with the accuracy required nowadays
in ab-initio calculations.
Furthermore, the centers of the Wannier functions are used to compute
the macroscopic polarization of Si and GaAs in zero electric field.
The effective
charges of GaAs, obtained by finite differentiation of the polarization,
agree with the results of linear response theory.
\end{abstract}
%]

\narrowtext

Since their introduction in 1937, the Wannier functions~\cite{Wannier} (WFs)
have played an important role in the theoretical study of the properties of
periodic solids.~\cite{Blount}
Very recently they have been one of the main
ingredients of a novel theory of the electronic polarization in
terms of a Berry phase.~\cite{Polarization}
Notwithstanding, in computational applications, the representation of the
electronic wavefunctions with Bloch orbitals is the method of choice.
In fact, for periodic systems, the Bloch theorem allows to exploit
the translational invariance of the solid, and to restrict the
problem to one unit cell. The properties of the infinite solid are
recovered with an integral over the Brillouin zone, which can be approximated
with a finite sum.

The WFs extend in principle all over the solid, and it is possible to
compute only approximate WFs which are constrained to be zero outside
a localization region (LR). In recent years, these approximate WFs turned
out to be a key-concept in the development of electronic-structure methods
whose computational cost scales linearly with the system
size.~\cite{GalliPar,Mauri1}
Furthermore these Wannier-like functions provide a very
promising way to study the electronic properties of a
solid in presence of a macroscopic electric field.~\cite{Elecfield}

For these reasons, it is now important to develop ab-initio methods to
compute the approximate WFs of a crystal, to study their properties as a
function of the LR, and to show in practice that using WFs it is possible
to extract structural and electronic properties of materials as accurate
as those obtained with Bloch functions.

The WFs are related to the Bloch functions by a unitary transformation.
However, this transformation is highly non-unique because
the Bloch functions are determined only up to a multiplicative
phase factor which introduces a large ambiguity in the localization
properties of the resulting WFs. For one-dimensional periodic solids,
with a finite gap, it has been shown analytically that an appropriate choice of
the
phases of the Bloch functions leads to WFs which decay exponentially in
space.~\cite{Wprop}
In a real solid, band-crossing and phase freedom make the problem of
localization particularly hard. A possible approach in this direction
has been recently discussed by Sporkman {\it et al}.~\cite{Sporkman}.
Although they obtained reasonably localized WFs also for fcc transition metals,
they did not
test the accuracy of their functions against any physical properties.

Several years ago, W. Kohn proposed to compute the WFs of solids
by minimizing the total energy in a variational scheme where
the trial functions were localized.~\cite{Kohn2}
However, from a computational point
of view the orthogonality constraint was a major problem and a few numerical
experiments have been performed along these lines.~\cite{Andreoni}
Nowadays, after several advances towards an order-N method for electronic
structure calculations, it is simpler to build an electronic structure code
entirely based on WFs. One example is presented in this work.

In the framework of Density Functional Theory (DFT),
Galli and Parrinello~\cite{GalliPar} introduced non-orthogonal localized
orbitals to minimize the total energy and to obtain the
electronic ground-state. In Ref~\onlinecite{Mauri1},
a total energy functional was proposed, which is minimized by
orthogonal orbitals and has the same minimum as the standard total
energy functional. Within a tight-binding formalism, this energy functional has
been used
to obtain almost orthogonal localized orbitals which reproduce
the ground state properties of silicon and carbon.

In this work we implement this functional in a self-consistent scheme
and we test extensively the practical possibility to describe
with Wannier functions the structural, electronic and dielectric properties of
materials at the level of accuracy obtained with Bloch wavefunctions.
Focusing on small unit cell systems we can
use very large LR and check the convergence of the physical
properties to the exact ground-state. At variance with the approach
of Ref.~\onlinecite{GalliPar,Mauri1} we use explicitly the translational
properties of the
WFs during the minimization of the total energy functional.~\cite{Elecfield}
At the minimum, the orbitals are almost orthogonal and are a good approximation
of a set of WFs for the system.
We study two crystals: silicon and gallium arsenide. We use DFT in the
local density approximation (LDA), and describe the atoms with norm-conserving
pseudopotentials. For each system we obtain the total energy, the lattice
constant, the bulk modulus and the frequency of the zone-center
transverse optical phonon and we study their dependence on
the size of the LR. LRs containing up to 342 atoms for Si and 216 atoms
GaAs are considered,
and our results are compared with converged values obtained
with a plane waves (PW) pseudopotential code based on Bloch orbitals.
We show that the error associated to the localization
can be made lower than the errors usually associated to the
use of LDA or pseudopotentials. We also address the ability of the approximate
WFs
to describe the macroscopic polarization of semiconductors in zero
electric field by computing the Born effective charges of GaAs.
We show that the effective charges extracted from our approximate WFs
are in good agreement with those obtained with a linear response approach
based on Bloch functions.
%
% METHOD
%

A system of $N$ interacting electrons described in the framework of DFT-LDA,
can be studied by introducing $N/2$ orbitals which describe an auxiliary
system of non-interacting electrons.~\cite{KS} In a periodic solid, the Bloch
theorem allows to label these states with a {\bf k}-vector in the
first Brillouin zone and a band index $n$. In insulators the number of
occupied bands is one half the number of electrons $N_{el}$ contained
in one unit cell. An equivalent representation can be obtained using
a set of WFs $|w_{l,n}\rangle$, where $n$ is the band index,
$l$ indicates the Bravais lattice vector ${\bf R}_l$. The WFs are orthonormal
and
$|w_{l,n}\rangle$ is obtained by translating the function centered at the
origin
by ${\bf R}_l$, i.e. $|w_{l,n}\rangle = \hat T_{{\bf R}_l} |w_{0,n}\rangle$.

The WFs are not unique: their shape and localization in real space
are arbitrary. However the
physical quantities computed from WFs do not depend on their
shape. In  particular the electronic contribution to the macroscopic
polarization per unit cell has a very compact expression in terms of
WFs,~\cite{Polarization} i.e.
$ {\bf P}_{el} = - 2 \sum_n \langle w_{0,n}|{\bf r}|w_{0,n}\rangle $.
This expression can be used to compute the Born effective charges
which are the derivative of the polarization with respect to atomic
displacements in zero electric field. Using WFs this quantity is
immediately available from a finite numerical differentiation.

In Ref.~\onlinecite{Mauri1} and \onlinecite{Elecfield} it has been shown
that WFs for a solid can be obtained directly by minimizing the following
functional:
\begin{eqnarray}
  E_{tot}\left[ \{ v \}, \eta \right]
  &=&
  \sum_{n} \sum_{l,m}
  2 Q^{0,l}_{n,m}
  \langle v_{0,n} |
   - \frac{1}{2} \nabla^{2} + \hat{V}_{NL}
   | v_{l,m} \rangle \nonumber \\
  &+& F\left[ \tilde{n} \right]
  + \eta ( N_{el} - \tilde{N} )
\label{energy}
\end{eqnarray}
where
$
 Q^{0,l}_{n,m} = 2\, \delta_{l,0} \, \delta_{n,m}
 - \langle v_{0,n} | v_{l,m} \rangle
$, $\tilde{N}$ is the integral over one unit cell of the
charge density $\tilde{n}({\bf r})$ defined as
\begin{equation}
 \tilde{n}({\bf r}) =
 \sum_{k,n} \sum_{l,m} 2
 Q^{k,l}_{n,m} \,
 \langle v_{k,n} | {\bf r} \rangle \langle {\bf r} | v_{l,m} \rangle,
 \label{charg}
\end{equation}
$\hat{V}_{NL}$ is the non-local part of the pseudopotential
and $F\left[ \tilde{n} \right]$ is the sum of the local, Hartree
and exchange-correlation energies.
$\eta$ is an energy parameter which is fixed in such a way to be higher
of the highest occupied eigenvalue.
The functions $| v_{l,n} \rangle$ are obtained by translating
$| v_{0,n} \rangle$ , i.e. $|v_{l,n}\rangle = \hat T_{{\bf R}_l}
| v_{0,n} \rangle$ and therefore they do not add any additional degrees of
freedom.
The charge $\tilde{n}({\bf r})$ is periodic in the
unit cell. Although no orthogonality constraint is explicitly imposed
on the $|v_{l,n}\rangle$, at the minimum, the $|v_{l,n}\rangle$ are orthonormal
and form
a set of WFs for the solid.~\cite{Mauri1}

In our calculation, we represent the functions $|v_{l,n}\rangle$
on a uniform cubic real-space mesh with spacing $h$ in each direction
${\bf r}_{ijk}=(i h,j h,k h)$, where $ i,j,k$ are integers.
Since it has been shown that the WFs of insulators can be chosen
exponentially localized, we impose $\langle {\bf r}_{ijk} \, | v_{0,n} \rangle$
to be zero
if ${\bf r}_{ijk}$ is outside a cubic region of size $2 a_{LR}$. The non-zero
coefficients
$\langle {\bf r}_{ijk} \, | v_{0,n} \rangle$ are obtained by minimizing the
total energy
Eq.~\ref{energy}. The imposition of localization is a variational approximation
for
the total energy which, at the minimum, gives orbitals which are not exactly
orthonormal.~\cite{Mauri1} By increasing the size of the localization region
the variational estimate of the energy improves and the deviation
of the orbitals from orthonormality is reduced. Therefore the orbitals converge
to a set of WFs for the system. Note that, if localization is imposed,
the sums over $l$ appearing in Eq.~\ref{energy} and $(k,l)$
in Eq.~\ref{charg} become finite and determined by the
set $(l,m)$ of LR that overlap with all the LR $(0,n)$ of the first unit cell.

In order to compute $E_{tot}\left[ \{ v \} , \eta \right]$ we need to
apply $-\frac{1}{2}\nabla^2 + \hat{V}_{NL} $
to $|v_{0,n}\rangle$. We evaluate these operators directly on the real space
grid.
For the non-local part of the Hamiltonian, we used the technique proposed by
King-Smith {\it et al}.~\cite{ks} to optimize the Kleinman-Bylander
projectors~\cite{Kleinman} for a real-space evaluation of matrix elements.
Thus,
for each atom in the position ${\bf \tau}_s$,
the Kleinman-Bylander projector is non-zero only on the mesh points contained
in a sphere
with radius $R_{cut}^{s}$ (core region).
For the kinetic energy operator we evaluate the Laplacian with a
finite differences formula as Chelikowsky {\it et al}.~\cite{cheli} which
delocalizes the orbital only up to $M$ points in each direction,
where $M$ is the degree of the expansion.
Once the size of the LRs $a_{LR}$ and of the core
regions $R_{cut}^{s}$ have been fixed,
$\langle {\bf r} | -\frac{1}{2}\nabla^2 + \hat{V}_{NL}
| v_{0,n} \rangle$ will be zero outside a cube with size
$2(a_{LR} + max\{ M h,2R^s_{cut} \})$. Since the charge $\tilde{n}({\bf r})$ is
periodic
we evaluate it on all the nodes of the real space mesh within one unit cell
and we compute the Hartree energy by solving the Poisson equation
in the unit cell with a Fast Fourier Transform (FFT).

%
%  RESULTS
%

We applied our approach to crystalline Si and GaAs.
The Bravais lattice is fcc and the unit cell contains 2 atoms.
The electronic structure is described by four occupied orbitals.
We considered four LRs centered on the bonds connecting one atom
with its four nearest neighbours. We use norm-conserving non-local
pseudopotentials which have been optimized with
$R_{cut}$ = 4 a.u.~\cite{notapseudo}.
Exchange and correlation effects are treated using the Perdew and Zunger's
parameterization.~\cite{PZexc}
The grid spacing has been chosen as $h = a/24$, where $a$ is the size of the
conventional cubic cell. This grid corresponds to a PW cut-off of
$54.0$ Ry and $49.8$ Ry for the density in the case of Si and GaAs respectively
if one uses the experimental lattice constant. For the kinetic energy we used a
very
conservative choice of $M$ which was set equal to $14$.
The free parameter $\eta$ was fixed to $3.0$ Ry and $4.0$ Ry for Si and GaAs
respectively. The values of these parameters are sufficient to give
energies converged within a few mRy for a given size of the LR.
We compare the results obtained with WF with those obtained using Bloch
orbitals.
In this case we expand the Bloch orbitals
in a PW basis with cut-off of 48 Ry for Si and 56 Ry for GaAs, and we use 28
special {\bf k}-points in the irreducible Brillouin zone.
With these parameters the error in the total energy is lower than $0.5$ mRy.

In Fig.~\ref{fig1} we show the convergence of the total energy of Si computed
with our method for different sizes of localization $a_{LR}$ and
compare the results with the exact value obtained by the
conventional diagonalization of the Hamiltonian using Bloch orbitals.
The figure shows that the error in the localization is 5 mRy
if the LR contains more than $216$ atoms.
In this case the error due to the localization is
comparable to the error introduced by the use of a real space grid.

In order to measure the deviation of the approximate WFs from orthonormality,
we can consider the quantity
$\Delta N = N_{el} - \tilde{N}= 2\sum_{l,m,n}(\delta_{l,0} \, \delta_{n,m}-
\langle v_{0,n} | v_{l,m} \rangle  )^2$.
In silicon,
increasing the size of the LRs from $a_{LR} = 14 h$ to
$a_{LR} = 43 h$, $\Delta N$ decreases
from $2.7 \times 10^{-3}$ to $6.3 \times 10^{-5}$.
In Fig.~\ref{fig2} we show the electronic charge density of silicon
computed along the $(111)$ direction. The charge is converged within
$0.1 $\% with $a_{LR} = 29\,h$.

In Table~\ref{table1} we show the total energy, the lattice constant, the bulk
modulus
and the frequency of the zone-center optical phonon as a function
$a_{LR}$ for silicon. The convergence of the theoretical lattice
constant is very rapid,
the error being less than $1 $\% with $a_{LR}=19\,h$.
With this size of the LR the computed bulk modulus is within $10$ kbar
from the converged value.
The frequency of the zone-center transverse optical phonon
is converged to $0.6 $\% for $a_{LR} = 24\,h$
(we used here the experimental lattice constant $a = 10.26$ a.u.).

In silicon the macroscopic polarization and its derivative with respect to
the atomic displacements, the Born effective charges are both zero.
We computed this polarization per unit cell using the equation:
\begin{equation}
{\bf P}_{tot}= -{2} \sum_{n} \sum_{l,m}
 Q^{0,l}_{n,m} \,
 \langle v_{0,n} | {\bf r} | v_{l,m} \rangle + {\bf P}_{ions},
 \label{polarizza}
\end{equation}
where the presence of the matrix $Q^{0,l}_{n,m}$
accounts for the approximate orthogonality of the orbitals.~\cite{Elecfield}
Here
${\bf P}_{ions}$ is the ionic contribution to the macroscopic polarization.
We have verified that the total polarization of Si is zero
(modulus a quantum equal to ${\bf R}_l$).
The accuracy of this zero depends on the degree of orthogonality of
the WFs. With our parameters, we find a value of
$| {\bf P}_{tot} | = 1.2\times10^{-2}$ a.u.
for $a_{LR} = 14\,h$ and $3.6\times10^{-4}$ a.u. for $a_{LR} = 43\,h$.

The results for GaAs are reported in Table~\ref{table2}.
Table~\ref{table2} shows a convergence with respect to $a_{LR}$ of the computed
physical
properties similar to that of Si.
GaAs is a polar semiconductor
with non-zero effective charges.
Using the WF we computed the effective charges by finite differentiation
of the macroscopic polarization with respect to the atomic displacements.
The convergence of the effective charges with respect to
the sizes of the LRs is shown in  Table~\ref{table2}, where for
comparison we also report the results obtained with Bloch orbitals
and linear response.~\cite{effetive_charges_linear_response}
We note that when the size of the LRs is equal to $29\,h$ the effective charges
obtained by displacing As or Ga are equal and opposite in sign within
$0.18$, and, in a linear response calculation, this accuracy is reached
with a Brillouin zone sampling of $10$ {\bf k}-points.

Finally, in Fig.~\ref{fig2} we show one example of Wannier-like
orbitals for GaAs along the $(1,1,1)$ direction.
The WFs corresponding to two different LRs are displayed. In both cases
the Wannier-like orbitals are well localized around the bond
center, and the center of each WF is displaced towards the arsenic atom.

In conclusion, we have presented a real-space, self-consistent
computation of the WFs of Si and GaAs. The scheme provides approximate
WFs which are constrained to be zero outside a cubic region.
We showed that it is possible, by using sufficiently large
LR, to extract from these WFs the structural and dynamical
properties of Si and GaAs, with an accuracy comparable to the
standard {\it ab-initio} methods. These results can have important
implications in the future developments of {\it ab-initio},
order-N methods based on Wannier orbitals.
Furthermore, we showed that the approximate WFs can give a
good estimate of the electronic polarization in zero electric
field and of its derivatives with respect to atomic displacements.
The ability of these approximate WFs to describe the
electronic structure of a solid in an external electric field is
currently investigated.

We gratefully acknowledge A. Pasquarello and R. Resta for useful discussions.
This work was supported by the Swiss National Science Foundation under
Grant No. 20-39528.93, by the US National Science
Foundation (NSF) under Grant No. DMR-9120269,
by the  Materials Sciences Division of the US
Department of Energy under Contract No. DE-AC03-76SF00098.

%
% FIGURES
%

\begin{figure}
\caption{Si: (a): Total energy versus the size $a_{LR}$ of the LRs.
(b): Convergence of the charge density along the
(111) direction. The lines correspond to a Fourier
Transform interpolation of the values of the charge density on the
nodes of the real-space mesh inside the $8$-atom cell (full circles).
$a_{exp}$ indicates the experimental lattice constant ($a_{exp} = 10.26$ a.u.).
}
\label{fig1}
\end{figure}
\begin{figure}
\caption{GaAs: Wannier function $\langle {\bf r} | w_{0,1} \rangle$ centered
on ${\bf b}_{0,1} = a/8\,(1,1,1)$ along the ${\bf r} = x (1,1,1)$ direction.
Full circles correspond to the values of this WF on
the nodes of the real-space mesh in this direction. The lines correspond to a
Fourier interpolation.
}
\label{fig2}
\end{figure}
%
% TABLES
%
%
\begin{table}
 \begin{tabular}[tb]{lcccc}
  $a_{LR}$ (h) & $\Delta E_{tot}$ (mRy) & $a_{0}$ (a.u.) & $B_{0}$ (kbar) &
  $\omega_{TO}$ (cm$^{-1}$)\\
 \tableline
   14 [8]  & 180.7 & 10.47 & 921 & 594 \\
   19 [26] &  65.7 & 10.29 & 944 & 543 \\
   24 [64] &  27.8 & 10.24 & 953 & 520 \\
   34 [216]&   5.4 & 10.21 & 951 &     \\
   43 [342]&   1.5 &       &     &     \\
  \hline
   $\infty$&     0 & 10.20 & 941 & 517
 \end{tabular}
\caption{Si: Error ($\Delta E_{tot} = E_{tot}(a_{LR}) - E_{tot}(\infty)$)
in the total energy, lattice constant ($a_{0}$), bulk modulus ($B_{0}$)
and frequency of the zone-center transverse optical phonon ($\omega_{TO}$)
for different sizes of localization ($a_{LR}$).
The numbers in square brackets  in the column
for ($a_{LR}$) correspond to the number of atoms inside each LR. The
data for $a_{LR} = \infty$ are the results obtained
with a standard PW code using Bloch orbitals.}
\label{table1}
\end{table}
\begin{table}
 \begin{tabular}[tb]{lcccccc}
  $a_{LR}$ (h) & $\Delta E_{tot}$ (mRy) & $a_{0}$ (a.u.) &
  $B_{0}$ (kbar) & $\omega_{TO}$ (cm$^{-1}$) & $Z^{\star}_{As}$ &
$Z^{\star}_{Ga}$ \\
 \tableline
   19 [26]    &    63.4  & 10.55 & 792 & 277 & 2.08 & $-$2.28 \\
   24 [64]    &    26.7  & 10.50 & 807 & 269 & 1.96 & $-$2.21 \\
   29 [126]   &    11.9  & 10.48 & 804 & 260 & 2.02 & $-$2.20 \\
   34 [216]   &    4.7   & 10.48 & 794 & 259 &      &         \\
 \hline
 $\infty$     &     0    & 10.48 & 760 & 268 & 2.17 & $-$2.34
 \end{tabular}
\caption{GaAs:
 Error ($\Delta E_{tot} = E_{tot}(a_{LR}) - E_{tot}(\infty)$) in the total
energy,
 lattice constant ($a_{0}$), bulk modulus ($B_{0}$) and frequency of the
zone-center
 optical phonon ($\omega_{TO}$) for different sizes of LRs $a_{LR}$.
 The numbers in in square brackets in the column for ($a_{LR}$) correspond to
 the number of atoms inside each LR.
 Born effective charges ($Z^{\star}_{As}$), ($Z^{\star}_{Ga}$) for As and Ga
atoms
 respectively.
 The data for $a_{LR} = \infty$ are the results obtained
 with a standard PW code using Bloch orbitals, and a linear response approach
for the
 effective charges.}
\label{table2}
\end{table}
\end{document}